\documentstyle[aps,epsfig]{revtex}
\begin{document}                                                              
\newcommand{\be}{\begin{eqnarray}}
\newcommand{\ee}{\end{eqnarray}}
\title{Two- and Three-photon Fusion in Relativistic Heavy Ion Collisions}
\bigskip
\author{C.A. Bertulani$^{a)}$,  and F. Navarra$^{b)}$} 
\address{
a) Department of Physics, Brookhaven National Laboratory,
Upton, New York 11973-5000, USA\\
b) Instituto de F\'\i sica, Universidade de S\~ao Paulo,
C.P. 66318, 05315-970 S\~ao Paulo, SP, Brazil\\
}
\maketitle
\bigskip
\begin{abstract} 
The production of mesons in ultra-peripheral collisions of relativistic heavy
ions is re-analyzed using a projection technique to calculate the amplitudes for the
appropriate Feynman diagrams. The virtuality of
the exchanged photons is fully accounted for in this approach. 
In the case of two-photon fusion, it is explicitly shown that the inclusion of 
nuclear form factors validates
the equivalent photon 
approximation. However, this does not apply to three-photon 
fusion cross sections. The cross section
of $J/\psi$ production in ultra-peripheral collisions at RHIC and LHC are
shown to be much smaller than the cross sections for the production
of $C=even$ mesons of similar masses.
\end{abstract}

\pacs{}

\section{Introduction}

Two-photon
physics is the dominant process in $e^+e^-$ colliders. This was
first shown by Brodsky, Kinoshita and Terazawa \cite{BKT70}.  
In an earlier paper, Low \cite{Low60}
showed that one can relate the particle production
by two real photons (with
energies $\omega_1$ and $\omega_2$, respectively)
to the particle's decay width, $\Gamma_{\gamma\gamma}$. Since both
processes involve the
same matrix elements, only the 
phase-space factors and polarization summations are
distinct. Low's formula is
\be
\sigma (\omega_1, \ \omega_2 ) = 8\pi^2 {\Gamma_{\gamma\gamma} \over M} \ 
\delta (4\omega_1 \omega_2 - M^2) \ ,
\label{photon}
\ee
where $M$ is the particle mass, $\Gamma_{\gamma\gamma}$ its decay
width, and the delta-function accounts for energy conservation.

Another important theoretical development was the realization that
the cross sections in colliders are well described by replacing 
the virtual photons by an equivalent field of real photons.
One often uses the concept of an equivalent photon number, 
$n(\omega)$, with energy $\omega$. This approximation, called the
Weizs\"acker-Williams method  \cite{Fe24} (or the equivalent
photon approximation) yields for the particle production
in colliders \cite{BKT70}
\be
\sigma = \int d\omega_1 d\omega_2
{n_1(\omega_1)\over \omega_1} \ {n_2(\omega_2)\over \omega_2}
\ \sigma_{\gamma\gamma} (\omega_1, \ \omega_2) 
\label{WW}
\ee

To our knowledge, ref. \cite{BB88} was the first to apply
a similar approach to study particle production 
in relativistic heavy ion collisions. As compared to 
$e^+e^-$ colliders heavy ions  carry the advantage
of a larger coupling constant ($Z\alpha$),
which increases the cross sections by a large factor. 
The disadvantage is that
one needs to separate the final products from those
created by strong
interaction processes. Inserting eq. \ref{photon}
in \ref{WW} and using the equivalent photon numbers
appropriate to heavy ions  the following expression
was obtained in ref. \cite{BB88}, to leading logarithmic order, 
\be
\sigma = {128\over 3} \ Z^4\alpha^2 {\Gamma_{\gamma\gamma}\over
M^3} \ \ln^3 \left( {2\gamma \delta \over M R} \right) \ ,
\label{ln}
\ee
where $\delta = 0.681..$, 
$\gamma$ is the Lorentz factor (e.g., $\gamma = 108$
for the RHIC collider at Brookhaven), and
$R$ is a parameter which depends on the mass of the
produced particle. If $M$ is much smaller than the inverse of 
a typical nuclear radius,  then $R = 1/M$, otherwise
$R$ is the nuclear radius.
These choices reflect the uncertainty relation in the
direction transverse to the beam, as explained in ref. \cite{BB88}. 
Since spin 1 particles cannot couple to two real photons
\cite{Ya50}, 
one expects that only spin 0 and spin 2 particles are produced.

Following these ideas, the two-photon fusion
mechanism in heavy ion collisions was exploited by several authors, 
including the possibility to search for the Higgs boson 
\cite{Pa89,Grab89,Dree89,BF90,CJ90,Nor90,Baltz98,Rold00}. At present, 
there are experiments at RHIC/Brookhaven, 
and proposed ones for the Large Hadron Collider at CERN (LHC) \cite{PEC},
which aim to study these phenomena.    
For mesons the cross section is very sensitive to
the minimum impact parameter, and refs. 
\cite{BF90,CJ90} have shown that corrections to
eq. \ref{ln} are substantial. 
These corrections are of geometrical nature and use the
equivalent photon method, as in eq. \ref{ln}.

Due to the large theoretical
and experimental interest in these phenomena  
\cite{Pa89,Grab89,Dree89,BF90,CJ90,Nor90,Baltz98,Rold00,PEC} 
(see also ref. \cite{BHT98} and
references therein), it is important to calculate the
production mechanism with an alternative approach.  
We use the projection method of ref. \cite{Nov78}
to obtain the meson-production amplitude in terms of
the amplitude for production of quark-anti-quark pairs
by the time-dependent field of the colliding nuclei.
In section 2 we start with a calculation for the production of
parapositronium in heavy ions colliders. This will define  
the calculational steps we need for the production of mesons.
In particular, we show that the results
agree with a recent calculation for this process \cite{serbo},
thus validating the projection method. 
In section 3
we extend the calculation to the production of $C=even$
mesons. In this case, one has to account for the nuclear form factors.
We show that the equivalent photon  method is obtained as a consequence
of the cutoff of large photon momenta, imposed by the inclusion of the
nuclear form factors.  
In section 4 we calculate the cross section for the production of 
vector
mesons ($C=odd$) by three-virtual photons. 
In particular, we show that the production rates
for  $J/\psi$ mesons are 
many orders of magnitude smaller than for the $C=even$ mesons of similar
masses.

\bigskip\bigskip\bigskip

\begin{center}
\begin{figure}[t]
\epsfig{file=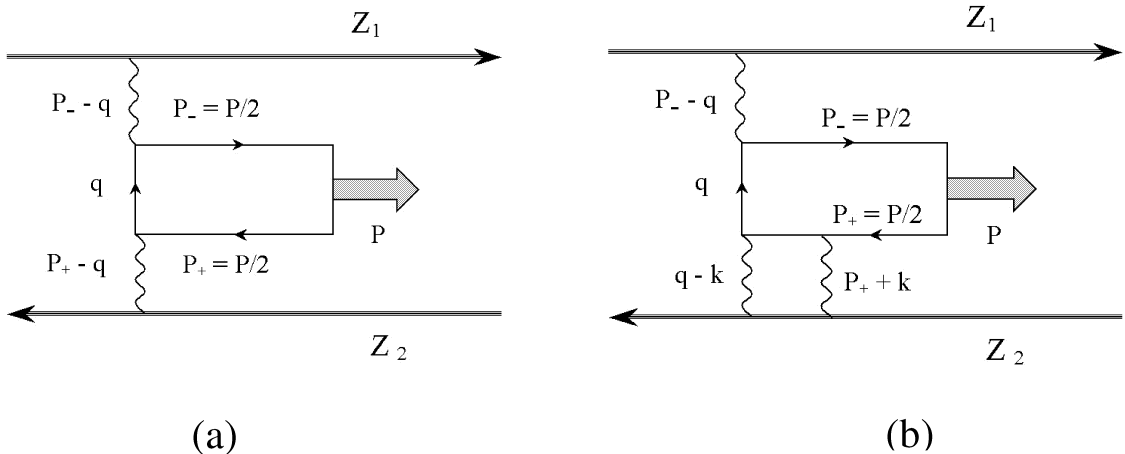,height=6.0cm}
\caption{\small
Feynman graphs for two- and three-photon fusion in ultra-peripheral
collisions of relativistic heavy ions.
}
\label{f1}
\end{figure}
\end{center}
\section{Two-photon fusion in heavy ion colliders}

In the laboratory frame the Fourier components of the classical electromagnetic field 
at a distance ${\bf b}/2$ of 
nucleus 1 with charge $Ze$ and velocity $\beta$, is given by (in our notation
$q = (q_0,{\bf q}_t, q_3)$, and $q_3 \equiv q_z$)
\be
A^{(1)}_0(q)=-8\pi^2Ze\ \delta(q_0-\beta q_3){e^{i{\bf q}_t .{\bf b}/2}
\over {q^2_t +q_3^2/\gamma^2}}\ \ \ \ \ \ \
{\rm and} \ \ \ \ A^{(1)}_3 = \beta A^{(1)}_0
\label{EMF}
\ee
For the field of nucleus 2, moving in the opposite direction, we replace $\beta$
by $-\beta$   and ${\bf b}$ by $-{\bf b}$ in the equations above. Although 
$\beta \simeq 1$ in relativistic colliders, it is important to keep 
them in the key places,
as some of their combinations will lead to important 
$\gamma = (1-\beta^2)^{-1/2}$ factors.

The matrix element for the production of positronium is 
directly obtained from the 
corresponding matrix element
for the production of a free pair (see fig. \ref{f1}(a)), with the requirement that
$P_+=P_-=P/2$, where $P$ is the momentum of the final bound state.
That is 
\be
{\cal M} &=& 
{\cal M}_1 +{\cal M}_2 \nonumber \\ 
&=& -i e^2 \bar{u}({P\over 2})\left[ \int{d^4q\over{(2\pi)^4}}
\not\!\! 
A^{(1)}({P\over 2}-q)
{{\not\! q + M/2}\over {q^2 - M^2/4}} 
\not\!\! A^{(2)}({P\over 2}+q) + 
\not\!\! A^{(1)}({P\over 2}+q)
{{\not\! q + M/2}\over {q^2 - M^2/4}} 
\not\!\! A^{(2)}({P\over 2}-q)
\right] v({P\over 2}) \ ,\label{M1} 
\ee
where $M$ is the positronium mass.

The treatment of bound states in quantum field theory is a very
complex subject (for reviews, see \cite{BYG85}). In our case,
we want to use the matrix element for free-pair production
and relate the results for the production of a bound-pair.  
A common trick used in this situation is
to convolute the matrix
element given above with the bound-state wave function. 
One can show (see, e.g., \cite{Nov78})
that this is equivalent to the use of a projection operator of the
form
\be
\bar{u} \cdots  v \longrightarrow  
\ {\Psi(0)\over 2\sqrt{M}}
\ {\rm tr} \left[\cdots (\not\!\! P + M) i \gamma^5\right] 
\ \ \ \ \ \ \ \ {\rm and} \ \ \ \ \ \ \bar{u} \cdots  v \longrightarrow 
\ {\Psi(0)\over 2\sqrt{M}}
\ {\rm tr} \left[\cdots (\not\!\! P + M) i \not\!\hat{e}^* \right] 
\label{traces}
\ee
where $\cdots$ is any matrix operator. The first equation applies 
to a spin 0 (parapositronium) and the second to spin 1 
(orthopositronium) particles,
respectively. In these equations $\Psi({\bf r})$ is 
the bound state wavefunction calculated
at the origin, and $\hat{e}^*$ is the 
polarization vector, given by
$\hat{e}^*_{\pm 1}= (0, 1/\sqrt{2}, \pm i/\sqrt{2}, 0)$ and
$\hat{e}^*_0 = (0,0,0,1)$.

Using eq. \ref{traces} in eq. \ref{M1}, one gets for the parapositronium production
\be
{\cal M}_1 &=& 4ie^2 \ {\Psi(0)\over 2\sqrt{M}} \ \int{d^4q\over{(2\pi)^4}}
{1 \over {q^2 - M^2/4}} \ \left[
\epsilon_{0\mu 3\nu} P_\nu q_\mu A^{(1)}_0 \left( P/2-q\right) \
A^{(2)}_3 \left( P/2+q\right)\right. \nonumber \\
&+&
\left. \epsilon_{3\mu 0\nu} P_\nu q_\mu A^{(1)}_3 \left( P/2-q\right) \
A^{(2)}_0 \left( P/2+q\right) \right] \ ,
\label{M1trace}
\ee
where $\epsilon_{\lambda \mu \nu \sigma }$ is the antisymmetric Levi-Civita tensor. 

Inserting the explicit form of the electromagnetic fields in eq. \ref{M1trace} we get
\be
{\cal M}_1 = ie^2 \ {\Psi(0)\over \sqrt{M}}
{2 \over (2\pi)^4} \ \left[
\epsilon_{0\mu 3\nu} P_\nu J_{0\mu 3} + \epsilon_{3\mu 0\nu} P_\nu J_{3\mu 0} 
\right]
 \ ,
\label{M1trace2}
\ee
where
\be
J_{0\mu 3}&=& \left( 8\pi^2 Ze\right)^2 \beta \int {d^4q\over q^2 - M^2/4}
q_\mu \ \delta\left[ \left(P/2-q\right)_0-\beta\left(P/2-q\right)_z\right]
\delta\left[ \left(P/2+q\right)_0+\beta\left(P/2+q\right)_z\right]
\nonumber \\
&\times& 
{\exp\left[i\left({\bf P}_t/2-{\bf q}_t\right) \cdot {\bf b}/2\right]
\over
\left[\left({\bf P}_t/2-{\bf q}_t\right)^2 +\left(P_z/2-q_z\right)^2/\gamma^2
\right]}\
{\exp\left[-i\left({\bf P}_t/2+{\bf q}_t\right) \cdot {\bf b}/2\right]
\over
\left[\left({\bf P}_t/2+{\bf q}_t\right)^2 +\left(P_z/2+q_z\right)^2/\gamma^2
\right]}
\ .
\label{J0mu}
\ee

The delta functions imply the conditions
\be
q_z = -P_0/2\beta \ \ \ \ \ {\rm and} \ \ \ \ q_0=-\beta P_z/2
\ .
\ee
We also note that $J_{0\mu 3} = - J_{3 \mu 0}$ and $\epsilon_{0\mu 3 \nu} = \epsilon_{3 \mu 0 \nu}$, and 
also that ${\cal M}_2$, which is obtained by the replacement $A^{(1)} \leftrightarrow A^{(2)}$ in ${\cal M}_1$,
is the same as ${\cal M}_1$, i.e., ${\cal M}_1={\cal M}_2$. 
In other words, the direct (fig. \ref{f1}(a)) and the exchange Feynman diagrams yield the
same result for the matrix element. This is a consequence of the
imposed condition that $P_-=P_+=P/2$ and of the projection onto the 
bound-state. It is an important result that will also show up in
the diagrams involving three-photons.
Gathering all these results, and using
$\epsilon_{0\mu 3 \nu} P_\nu I_\mu=|{\bf P} \times {\bf I}|$, we get
\be
{\cal M}=16 i {\Psi(0)\over \sqrt{M}} \left( Z\alpha\right)^2 \Big|{\bf P \times I}\Big| \ ,
\ee
where
\be
{\bf I} = \int {d^2q_t \ {\bf q}_t \over q_t^2+{\cal Q}^2} \
{1 
\over
\left[\left({\bf P}_t/2+{\bf q}_t\right)^2 +\omega_1^2/\gamma^2
\right]}\
{1 
\over
\left[\left({\bf P}_t/2-{\bf q}_t\right)^2 +\omega_2^2/\gamma^2
\right]}
\ ,
\ee
with
\be
{\cal Q}^2 = {M^2\over 2}+{P_t^2 \over 4} +{P_z^2\over 2 \gamma^2}
\simeq {M^2\over 2}+{P_t^2 \over 4}
\label{calQ}
\ee
and
\be
\omega_1 = {E/\beta - P_z\over 2} \ ,\ \  \
\omega_2 = {E/\beta + P_z\over 2} \ \ \ {\rm and}
\ \ \ \ 4\omega_1\omega_2 = M^2+P_t^2-P_z/\gamma^2 \simeq M^2+P_t^2
\ ,
\label{omega12}
\ee
where $E\equiv P_0$ is the total positronium energy.

We see that $\omega_1$ and $\omega_2$ play the role of the (real) photon
energies. For real photons one expects $4\omega_1\omega_2 = E^2$, as
in eq. \ref{photon}.

The two-photon fusion cross sections can be obtained   by using

\be 
d\sigma = \sum_{\mu} \left[
\int d^2b \ |{\cal M}(\mu)|^2\right]\  {d^3P\over (2\pi)^3 \ 2 E}
\label{sig}
\ee 

Since the important impact parameters for the
production of the positronium will be
$b > 1/m_e \gg R$, where $R$ is the nuclear radius,
the integral over impact parameter can start
from $b=0$. Thus, the integral over impact parameter 
in eq. \ref{sig} yields the delta function
\be
{1\over (2\pi)^2} \ \int \exp \left[i({\bf q}_t-{\bf q}'_t)
\cdot {\bf b} \right] \ d^2 b = \delta   
\left({\bf q}_t-{\bf q}'_t \right)
\ .
\label{delta}
\ee
We thus obtain
\be
{d\sigma \over
d^3 P} = {64 \over \pi ME} \ \Big|\Psi(0)\Big|^2\
 \left(Z\alpha\right)^4 \
\int {d^2q_t \ \over 
\left(q_t^2+{\cal Q}^2\right)^2} \
{({\bf P}_t\times {\bf q}_t)^2
\over
\left[\left({\bf P}_t/2+{\bf q}_t\right)^2 +\omega_1^2/\gamma^2
\right]^2}\
{1 
\over
\left[\left({\bf P}_t/2-{\bf q}_t\right)^2 +\omega_2^2/\gamma^2
\right]^2}
\ . \label{cross1}
\ee

We now show that the above equation is equal to the equation
obtained in ref. \cite{serbo}. 
First we change the variables to
\be
{\bf q}_{1t} = {{\bf P}_t\over 2}- {\bf q}_t \ ,
\ \ \ 
{\bf q}_{2t} = {{\bf P}_t\over 2}+ {\bf q}_t \ ,
\ \ \ \ q_{1z}= {P_z/2 - q_z \over \gamma} 
\  , \ \ \ \
q_{2z}= {P_z/2 + q_z \over \gamma } \ .
\label{qs}
\ee
It is easy to show that
\be
\left({\bf P}_t/2-{\bf q}_t\right)^2 +\omega_2^2/\gamma^2
= q_{1t}^2 +q_{1z}^2 = -q_1^2 \ , \ \ \ \ \
{\rm and} \ \ \ \ \ \
\left({\bf P}_t/2-{\bf q}_t\right)^2 +\omega_1^2/\gamma^2
= q_{2t}^2 +q_{2z}^2 = -q_2^2 \ ,
\ee
and that
\be
{\bf P}_t \times {\bf q}_t = {\bf q}_{1t} \times {\bf q}_{2t}
\ , \ \ \ \ {\rm and} \ \ \ \ \
q_t^2+{\cal Q}^2 \simeq q_t^2 - {M^2\over 2} + {P_t^2\over 4}
= q_1^2+q_2^2-M^2
\ee

The positronium wavefunction at the origin is very well known. It is
given by $ \Big|\Psi(0)\Big|^2 = M^3 \alpha^3/64\pi$, where $M$ is
the positronium mass.
Thus, eq. \ref{cross1} becomes
\be
E {d\sigma \over
d^3 P} ={\zeta (3) \over
\pi} \ {\sigma_0 \over M^2} \ J_B
\ ,
\label{sigJB}
\ee
where
\be
J_B = {M^2 \over \pi} \int {\bf A}^2 \delta
\left( {\bf q}_{1t}+ {\bf q}_{2t}-{\bf P}_t\right) d{\bf q}_{1t}
d{\bf q}_{2t} \ ,
\label{JB}
\ee
with
\be
{\bf A} = {{\bf q}_{1t}\times {\bf q}_{2t} \over
q_1^2q_2^2} \ {M^2 \over M^2 -q_1^2 -q_2^2} \ ,\ \ \ \ \
\zeta(3) = 1.202... \ ,  \ \ \ \ 
{\rm and} \ \ \ \ \ \sigma_0 = {4Z^4\alpha^7\over M^2} \ .
\ee
We have included the zeta-function $\zeta(3)$ to take into account
the production of the para-positronium in higher orbits, besides
the production in the K-shell. 

The equation \ref{sigJB} is exactly the same as 
eq. 2.23 of ref. \cite{serbo}.
Thus we have shown that the approach used in this article for
the production of a bound particle (in this case, the para-positronium)
by means of the two-photon fusion yields the same results as
in the approach of ref. \cite{serbo}. In that article the
total cross section for the production of the para-positronium was obtained
by separating the regions where a leading order logarithmic approximation
could be used and a region where the integral in eq. \ref{JB} 
could be solved
numerically. To verify their results we will follow a different route.
Using eq. \ref{cross1} we can do the integration over the angle
$\phi$ between ${\bf P}_t$ and ${\bf q}_t$ analytically. We get
\be
\sigma = {M^4\over 2\pi} \zeta(3) \sigma_0 \ \int dP_t dP_z dq_t {q_t^3 P_t^3
\over E} \ {N(q_t,P_t,P_z) \over  
\left(q_t^2+{\cal Q}^2\right)^2}
\label{triplei}
\ ,
\ee
where 
\be
N(q_t,P_t,P_z) = {2\pi \over b^4}
\ {\sqrt{a_2^2-1}(a_1^2-a_1a_2-2) +\sqrt{a_1^2-1}(a_2^2-a_1a_2-2)
\over
\sqrt{a_1^2-1}\sqrt{a_2^2-1}
(a1+a2)^3}
\ ,
\ee
where
\be
b = q_tP_t \ , \ \ \ \ 
a_1 = {P_t^2/4 +q_t^2 +\omega_1^2/\gamma^2 \over q_tP_t}
\ \ \ \  {\rm and} \ \ \ \ \ \
a_2 = {P_t^2/4 +q_t^2 +\omega_2^2/\gamma^2 \over q_tP_t}
\ .
\ee
The triple integral in eq. \ref{triplei} can be calculated numerically. For RHIC,
using $\gamma = 108$ and $Au+Au$ collisions, we find 
$\sigma = 19.4$ mb.  
For the LHC ,
using $\gamma = 3000$ and $Pb+Pb$ collisions, we find  
$\sigma = 116$ mb. These are in good agreement with the results
of ref. \cite{serbo}.
  
\section{ Production of $C=even$ Mesons}

We can extend the calculation of the previous section to account for
the production of mesons with spin $J = 0$ and $J =2$ by the two-photon
fusion mechanism.
The following procedure is to be adopted:
\begin{enumerate}
\item Replace the electron-positron lines by quark-antiquarks in the
diagram of figure 1(a).
\item $M$ in the following formulas will refer to the meson mass.
\item Replace $\alpha^2$ by $\alpha^2 \ (2J+1) \ 3 \sum_i Q_i^4$, where
3 accounts for the number of colors, and $Q_i$ is the fractional quark charge.
These two last factors will cancel out when we express
$\Big|\Psi(0)\Big|^2$ in terms of $\Gamma_{\gamma\gamma}$, the decay-width
of the meson. 
To understand how this is done, lets discuss the
basics of the annihilation process of a positronium
(see also ref. \cite{BLP82}). With
probability $\alpha^2$ the $e^-$ 
can fluctuate and emit a virtual photon with energy
$~ m_e$.
The electron recoils and can travel up to a distance $\sim 1/m_e$ 
(or time $\sim m_e$) to meet the
positron and annihilate. 
This occurs when $e^-$ and $e^+$ are both found close
together in a volume of size $(1/m_e)^3$, i.e., 
with a probability given by $|\Psi(0)|^2/m_e^3$.
Thus, the annihilation probability per unit
time (decay width) is  
$\Gamma \sim \alpha^2 |\Psi(0)|^2/m_e^2$. Angular momentum conservation
and CP invariance does not allow the ortho-positronium to decay into an even
number of photons \cite{Ya50}. 
The description of the annihilation process given above is 
thus only appropriate for the para-positronium. 
A detailed QED calculation yields
an extra  $4\pi$ in the formula above. This 
yields $\Gamma_{\gamma\gamma} (^1S_0)=
8.03 \times 10^9 \ s^{-1}$, while the experimental value \cite{Th70} is 
$7.99(11) \times 10^9 \ s^{-1}$, in good agreement with the
theory.
For mesons, including the color and the charge factors, as described 
before, the relationship between $\Psi(0)$ and  $\Gamma_{\gamma\gamma}$
arise due to the same reasons. One gets 
$\Gamma_{\gamma\gamma}=16\pi\alpha^2 \Big|\Psi(0)\Big|^2/M^2 \cdot 3 
\sum_i Q_i^4$. 

According to these arguments the connection
between $\Gamma_{\gamma\gamma}$ and $\Big|\Psi(0)\Big|^2$, extended to
meson decays,
should be valid for
large quark masses so that $1/m_q \ll \ \sqrt{<r^2>}$, where 
$\sqrt{<r^2>}$ is the mean size of 
the meson. Thus, it should work well for, e.g. charmonium states, 
$c\bar c$. In fact, Appelquist and Politzer \cite{AP75} have generalized
this derivation for the hadronic decay of heavy quark states, which besides
other phase-space considerations amounts 
in changing $\alpha$ to $\alpha_s$, the strong coupling constant.
This can be simply viewed as a way to get a constraint on the
wavefunction $|\Psi(0)|^2$ \cite{RW67}.  
One expects that these arguments
are valid to zeroth order in quantum
chromodynamics and in addition one should include relativistic corrections.
But, as shown in \cite{BLP82}, the inclusion of relativistic effects, 
summing diagrams to higher order in the perturbation series, is equivalent
to solving the non-relativistic Schr\"odinger equation.

\item Change the integration variable to ${\bf q}_{1t}$ and
${\bf q}_{2t}$.

\item Introduce form factors $F(q_{1t})$ and $F(q_{2t})$ to account for the
nuclear dimensions. This is a simple way to eliminate the integral
over impact parameters and will be justified 'a posteriori', i.e., when
we compare our results with those from other methods.
These form factors will impose a cutoff in $q_{1t}$ and $q_{2t}$, so that
\be
q_{1t}\ , \ \ q_{2t} \simeq {1\over R} \ll M
\label{llM}
\ ,
\ee
where $R$ is a typical nuclear size. Taking $R = 6.5$ fm, we get $1/R \sim 30$
MeV. This is much smaller than the meson masses.
As an outcome of this condition, we can replace ${\cal Q}^2 \sim M^2 /2$
in eq. \ref{calQ}.

\end{enumerate}

According the procedures 1-5 we get from eq. \ref{cross1},
\be
{d\sigma \over d P_z} = {16 (2J+1)\over \pi^2} {Z^4 \alpha^2 \over M^3}
\ \Gamma_{\gamma\gamma} \ {1\over E} 
\int d{\bf q}_{1t} d{\bf q}_{2t} \
({\bf q}_{1t}\times {\bf q}_{2t})^2\ 
{\left[ F_1(q_{1t}^2)F_2(q_{2t}^2)\right]^2 \over 
\left(q_{1t}^2+\omega_1^2/\gamma^2\right)^2
\left(q_{2t}^2+\omega_2^2/\gamma^2\right)^2}
\label{meson2}
\ee

Using eqs. \ref{omega12} we have
\be
E = \omega_1 + \omega_2 \ , \ \ \ \ 
\omega_1 - \omega_2 = P_z \ , \ \ \ \ \ 
{\rm and} \ \ \ \ \
\omega_1 \omega_2 = M^2/4
\nonumber
\ee
so that
\be
dP_z = \left( 1 + {M^2 \over 4 \omega_1^2} \right) d\omega_1
\ , \ \ \ \ \ {\rm and} \ \ \ \ \
E =  {\omega_1^2 +M^2/4 \over \omega_1}
\ .
\ee
Thus,
\be
{d\sigma \over d \omega_1}  =  \sigma^{(+)} \ {d{\cal N}_{2\gamma}(\omega_1)
\over d\omega_1}
= \sigma^{(+)}
 \ {1\over \omega_1}
\
n_1(\omega_1) n_2(\omega_2) \ ,
\label{epaf1} 
\ee
where
\be
\sigma^{(+)} = 8 \pi^2 (2J+1) {\Gamma_{\gamma\gamma} \over M^3}
\ \ \ \ {\rm and} \ \ \ \ \
n_i(\omega_i) =  {2\over \pi}\ Z^2 \alpha \ \int {dq \ q^3
\ \left[ F_i(q^2) \right]^2 \over 
\left(q^2+\omega_i^2/\gamma^2\right)^2} \ .
\label{epaf2}
\ee
We notice that $n(\omega)$ is the frequently used form
of the equivalent photon number which enters eq.
\ref{WW}. Thus, eqs. \ref{epaf1} and \ref{epaf2} are the result one 
expects by using the
equivalent photon method, i.e., by using eqs. \ref{photon} and
\ref{WW}. This is an important result, since it shows that the projection
method to calculate the two-photon production of mesons works even for
light quark masses (i.e., for $\pi^0$). In this case there seems to be no
justification for replacing the quark masses and momenta by half the
meson masses and momenta, as we did for the derivation of eq. 
\ref{meson2}. This looks quite intriguing, but it is easy to see that
the step 3 in our list of procedures adopted is
solely dependent on the meson mass, not on the quark masses, i.e., if they
are constituent, sea quarks, etc. Moreover, the projection method
eliminates the reference to quark masses in the momentum integrals. 
The condition \ref{llM} finishes the job, by eliminating the photon
virtualities and yielding the same result one would get with the
equivalent photon approximation.

In the next section we will extend
this approach to the calculation of vector meson ($J=1^-$) production
by three  photons. There we will also apply the results to light 
quark masses, but we will not be able to check the results against
the equivalent  photon method since we cannot calculate the process
as originating from the collisions of three real photons.

We now define a ``two-photon equivalent number'', 
${\cal N}_{2\gamma}(M^2)$, so that 
$\sigma = \sigma^{(+)} {\cal N}_{2\gamma}(M^2)$, where
\be
 {\cal N}_{2\gamma}(M^2) =\int d\omega {d{\cal N}_{2\gamma} \over d \omega}=
\int {d\omega\over \omega} \ n_1(\omega)
\ n_2\left({M^2\over 4\omega}\right) \ .
\ee 

To calculate this integral we need the equivalent photon numbers
given by eq. \ref{epaf2}. 
The simplest form factor one can use for this purpose
is the `sharp-cutoff' model,  which assumes that
\be
F(q^2) = 1 \ \ {\rm for} \ \ 
q^2 < 1/R \ ,
\ \ \ {\rm and }\ \ \ F(q^2) = \ 0 \ ,
\ \ {\rm otherwise} \ .
\ee

In this case, we can use the integral
\be
\int_0^{1/R} {dq\  q^3 \over \left( q^2 + \omega^2 /\gamma^2 \right)^2}
= {1\over 2}\ \left[
\ln \left( 1+{\gamma^2 \over \omega^2 R^2} \right)-
{1\over 1+ \omega^2 R^2/\gamma^2}\right]
\label{LLspec1}
\  ,
\ee
and get for the differential cross section
\be
{d\sigma \over d \omega} = \sigma^{(+)} \ {Z^4\alpha^2\over \pi^2\omega}
\ \left[
\ln \left( 1+{\gamma^2 \over \omega^2 R^2} \right)-
{1\over 1+\omega^2 R^2/\gamma^2}\right]
\
\left[
\ln \left( 1+{16\gamma^2  \omega^2 \over M^4R^2} \right)-
{1\over 1+  M^4R^2/ 16\gamma^2  \omega^2  }\right]
\label{LLspec2}
\ .
\ee
The spectrum possesses a characteristic $1/\omega$ dependence, except
for $\omega \gg \gamma/R$, when it decreases 
as $1/\omega^5$.

When the condition $\gamma \gg M R$ is met, we can neglect
the unity factors inside the logarithm in eq. \ref{LLspec1}, as
well as the second term inside brackets.
Then, doing the integration of \ref{LLspec2}
from $\omega = M^2R/4/\gamma$ to $\omega = \gamma/R$,
we get eq. \ref{ln}.
But, eq. \ref{LLspec2} is an improvement over eq. \ref{ln}.
Eq. \ref{ln} is often used in the literature, but it is
only valid for $\gamma \gg MR$. This relation does not
apply to, e.g., the Higgs boson production ($M_{H^0} \sim 100$
GeV), as considered in ref. \cite{Pa89}.

For quantitative predictions we should use a more realistic
form factor. The Woods-Saxon distribution, with 
central density $\rho_0$, size $R$, and
diffuseness $a$ gives a good description of the densities of
the nuclei involved in the calculation. 
However, this distribution is very well described by the convolution
of a hard sphere and an Yukawa function \cite{DN76}. In this case,
the form factors can be calculated analytically 
\be
F(q^2) =
{4\pi \rho_0 \over q^3} \  
\left[ \sin (qR) - qR \cos (qR) \right]
\ \left[ {1\over 1+q^2a^2}\right]
\label{formfact}
\ .
\ee
For $Au$ we use $R = 6.38$ fm, and $a = 0.535$ fm, with $\rho_0$
normalized so that $\int d^3 r \rho (r) = 197$. For $Pb$ the appropriate 
numbers are 6.63 fm, 0.549 fm, and 208, respectively \cite{Ja74}.
With this form factor the two-photon 
equivalent photon number $d{\cal N}_{2\gamma}/d\omega$ 
is also obtainable in a closed form. In table 1 we
show the cross sections for the
production of $C=even$ mesons at RHIC ($Au+Au$) and LHC ($Pb+Pb$)
using the formalism described above.

\bigskip

\begin{center}
Table 1. {\small Cross sections for two-photon production of ($C=even$) mesons 
at RHIC
($Au+Au$) and at LHC ($Pb+Pb$).}

\bigskip
\begin{tabular}{l|c|c|c|c|c|c|c|c|r}
\hline
& meson     & mass [MeV] & $\Gamma_{\gamma\gamma}$ [keV]& 
$\sigma^{(+)}$ [nb]
&${\cal N}_{2\gamma}^{RHIC}/10^3$& ${\cal N}_{2\gamma}^{LHC}/10^7$
&$\sigma^{RHIC}$ [$\mu$b] & $\sigma^{LHC}$ [mb] \cr      
\hline
& $\pi_0$   &  134  & $7.8 \times 10^{-3}$&99&  49     & 2.8 &  4940  & 28    \cr
& $\eta$    &   547  & 0.46     &   86 &       12      & 1.8 &  1000  &  16 \cr
& $\eta'$   &   958  & 4.2     &   147 &         5.1      & 1.4 &  746  &  21 \cr
& $f_2 (1270)$& 1275 & 2.4     &  179 &          3.0      & 1.2 &  544  & 22 \cr
& $a_2 (1320)$& 1318  & 1.0     & 67  &        2.9      & 1.1 &  195  & 8.2   \cr
& $\eta_c$  &   2981 & 7.5     &  8.7   &        0.38     & 0.7&  3.3 &  0.61  \cr
& $\chi_{0c}$& 3415  & 3.3     &   2.6   &      0.24     & 0.63 &  0.63 & 0.16  \cr
& $\chi_{2c}$&   3556  & 0.8     &  2.8 &       0.21     & 0.56&  0.59 & 0.15 \cr 
\hline
\end{tabular}

\end{center}

\bigskip

As pointed out in refs. \cite{BF90,CJ90}, one can improve the 
(classical) calculation
of the two-photon luminosities by introducing a geometrical factor 
(the $\Theta$-function in ref. \cite{BF90}), which affects the
angular part of the integration over impact parameters. 
This factor takes care of the
position where the meson is produced in the space surrounding
the nuclei. In our approach the form factors also introduce a geometrical
cutoff implying that the mesons cannot be produced inside the nuclei.
However, it is not easy to compare both approaches directly, as we obtain
a momentum representation of the amplitudes when we perform  
the integration over impact parameters to obtain eq. \ref{meson2}.
But we can compare the effects of geometry in both cases by using equation  
\ref{LLspec2}. After performing the integral over $\omega$,
we can rewrite it as
\be
\sigma = \int ds  {\cal L} (s) \sigma_{\gamma\gamma} (s) \ ,
\label{sigfin}
\ee
where $s = 4\omega_1\omega_2$ is the square of the center-of-mass
energy of the two photons, $\sigma(s)$ is given by eq. \ref{photon},
and ${\cal L}(s)$ is the ``photon-photon luminosity'', given by
\be
{\cal L}(s) = {1\over s} \ {Z^4\alpha^2\over \pi^2}
\int {d\omega\over \omega} \left[
\ln \left( 1+{\gamma^2 \over \omega^2 R^2} \right)-
{1\over 1+\omega^2 R^2/\gamma^2}\right]
\
\left[
\ln \left( 1+{16\gamma^2  \omega^2 \over s^2R^2} \right)-
{1\over 1+  s^2R^2/ 16\gamma^2  \omega^2  }\right]
\label{Luminosity}
\ .
\ee

\bigskip\bigskip

\begin{center}
\begin{figure}[b]
\epsfig{file=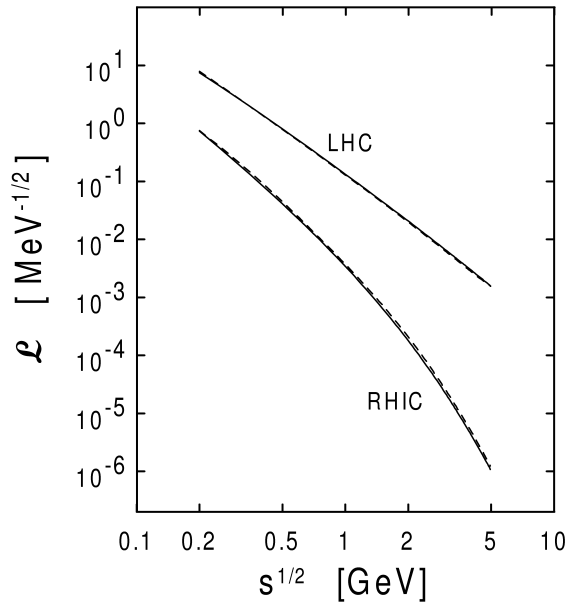,height=8.0cm}
\caption{\small
Two-photon luminosities (see definition in eq. \ref{sigfin}) at RHIC and
LHC. Dashed lines include a geometric correction.
}
\label{f3}
\end{figure}
\end{center}

In figure 2 we compare the result obtained by eq. \ref{Luminosity}
and that of ref. \cite{BF90}. 
The luminosities for RHIC ($Au+Au$)
and for LHC ($Pb+Pb$) are presented.
For RHIC the difference between the two 
results can reach 10\% for very large meson masses (e.g. the Higgs), but 
we notice that for the LHC the two results are practically identical, the difference 
being of the order of 3\%, or less, even for the Higgs.
Thus, the improved version
of eq. \ref{ln}, given by integrating eq. \ref{LLspec2}, is  
accurate enough to
describe meson production by two-photon fusion. Other effects, like
the interference between the electromagnetic and the strong 
interaction production mechanism in grazing collisions, must yield
larger corrections to the (non-disruptive) meson  production cross sections 
than a more elaborate description of geometrical effects.  

The results in this section are very important for our purpose of
calculating the production of vector mesons by means of three-photon
fusion in peripheral collisions. This could be a relevant process, e.g., 
for a study of the three-photon-vertex in charmonium production.

One might think that the calculation could be performed by
using the equivalent photon approximation that, as we have seen
in this section, works so well for $C=even$ mesons. 
However, the introduction of a third photon
leads to an additional integration, which implies that at least two
of the exchanged photons cannot be treated as real ones. 
Nonetheless, the results of this section paves the way to the 
calculation of production of
$C=odd$ mesons. Although the use of the
projection technique to systems composed of light quarks is
questionable, we have seen that it works, basically because of
the relation \ref{llM}, due to the inclusion of the
nuclear form-factors.

\section{Production of vector mesons}

Lets now consider the diagram of figure 1(b), appropriate for the
fusion of three photons into a $C=odd$ particle.
According to the Feynman rules, the matrix element
for it is given by
\be
{\cal M}_a = 
e^3 \bar{u}({P\over 2})\int{d^4q\over{(2\pi)^4}}
\int{d^4k\over{(2\pi)^4}}
\not\!\! 
A^{(1)}({P\over 2}-q)
{{\not\! q + M/2}\over {q^2 - M^2/4}}
\not\!\! A^{(2)}(q-k) 
{{\not\! k + M/2}\over {k^2 - M^2/4}} 
\not\!\! A^{(2)}({P\over 2}+k)  
 v({P\over 2}) \ .\label{M31} 
\ee
There will be 12 diagrams like this. But, as
we will see below, the upper photon leg in diagram of fig. 1(b)
can be treated as a real photon, meaning that the equivalent
photon approximation is valid for this piece of  the diagram. 

One has to use the second of the
eqs. \ref{traces} to account for the projection onto $C=odd$ particles. The
calculation of the traces is quite lengthy and was performed using the 
program FORM \cite{form}. 
We have found out that the particle is produced with its polarization
vector in the transverse direction, as the coefficients
accompanying $\hat{e}_0^*$ are of higher order in
$1/\gamma$. Neglecting such terms we get
\be
&&A^{(1)}_\alpha A^{(2)}_\beta A^{(2)}_\lambda
Tr[\gamma^\alpha (\gamma^\mu q_\mu+M/2)\gamma^\beta
(\gamma^\nu k_\nu+M/2)\gamma^\lambda (\gamma^\rho P_\rho + M)\hat{e}^*_\eta
\gamma^\eta] \nonumber \\
&=& 
- 16M  A^{(1)}_0 A^{(2)}_0 A^{(2)}_0
(k_0+\beta k_3) ({\bf q}_t - {\bf P}_t/ 2) \cdot \hat{\bf e}^*
\ .
\ee
The above product of the longitudinal components of $A_\mu$ yields
factors proportional to the delta-function, i.e.,
\be
 A^{(1)}_0 A^{(2)}_0 A^{(2)}_0
\propto
\delta \left[
{E\over 2} - q_0 - \beta ({P_3 \over 2} - q_3)
\right]
\delta \left[
 q_0 - k_0+\beta (q_3 - k_3)
\right]
\delta \left[
{E\over 2} + k_0 - \beta ({P_3 \over 2} + k_3)
\right]
\ .
\ee
These delta-functions lead to the conditions
\be
q_3 = -  {E\over 2 \beta} \ , \ \ \ 
q_0 = - {P_3\beta \over 2}
 \  ,
\ \ \ {\rm and}
\ \ \ 
k_0 = -\beta k_3 - \beta\omega_2 \ ,
\ee
where $\omega_2$ is given by \ref{omega12}.

The integral over $q_0$, $k_0$ and $k_3$ yields a factor 1/2, and the
matrix element becomes
\be
{\cal M}_a &=& {8\over \pi^2} (Z\alpha)^3 
\ \sqrt{M} \Psi(0)  \int d^2 q_t d^3 k
\ {({\bf q}_t - {\bf P}_t /2) \cdot \hat{\bf e}^*
\over q^2 - M^2/4}
\ {\omega_2 \over k^2 - M^2/4}\
\left[\left( {\bf P}_t/2 - {\bf q}_t\right)^2
+ \omega_2^2/\gamma^2 \right]^{-1}\nonumber \\
&\times&\left[\left( {\bf q}_t - {\bf k}_t\right)^2
 + \left( E/2\beta+k_3 \right)^2/\gamma^2 \right]^{-1}
\
\left[\left( {\bf P}_t/2 + {\bf k}_t\right)^2
+ \left( P_3/2+k_3 \right)^2/\gamma^2 \right]^{-1}\
\exp\left({{\bf q}_t \cdot {\bf b}/2}\right) \
\label{three1}
\ee

As in eq. \ref{cross1} we know that the nuclear form factors imply
\be
q_{t}\ , \ \ P_{t} \ , \ \  k_{t} \simeq {1\over R} \ll M
\ ,
\ee

\bigskip\bigskip\bigskip

\begin{center}
\begin{figure}[b]
\epsfig{file=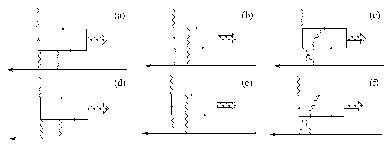,height=6.0cm}
\caption{\small
Feynman graphs for three-photon fusion in ultra-peripheral
collisions of relativistic heavy ions.
}
\label{f2}
\end{figure}
\end{center}

We can thus use the approximation for the propagator $(q^2-M^2/4)^{-1} 
\simeq -2/M^2$. 
Changing the integration variable from ${\bf k}$ to ${\bf k}'={\bf k} 
+ {\bf P}/2$, the propagators in the second line of \ref{three1} become
\be
\left[\left( {\bf q}_t +{\bf P}_t/2 - {\bf k}'_t\right)^2
 + \left( \beta E/2 - P_3/2 + k'_3 \right)^2/\gamma^2 \right]^{-1}
\
\left[ k^{'2}_t +k_3^{'2}/\gamma^2\right]^{-2}\
\ee
The integrand peaks sharply at ${\bf k}'_t=0$ and one can 
eliminate ${\bf k}'_t$
from the first term inside brackets.

Thus, the matrix element \ref{three1} becomes
\be
{\cal M}_a &=& -{16\over \pi}\ (Z\alpha)^3 
\ {\Psi(0)\over M^{3/2}}  \int d^2 q_t d k_t \ k_t 
\ {
({\bf q}_{1t} \cdot {\bf e}^*)
\exp\left( {{\bf q}_t \cdot {\bf b}/2} \right)
\over 
\left[ q_{1t}^2+\omega_2^2/\gamma^2\right]
} \nonumber\\
&\times&
\int dk_3 \left[ k_t^2+k_3^2/\gamma^2 \right]^{-1}
\left[ q_{2t}^2+(\omega_1+k_3)^2/\gamma^2\right]^{-1}
\left[ k_3+\omega_2-\omega_1-k_3^2/2\omega_2\gamma^2\right]^{-1}
\label{three2}
\ee
where ${\bf q}_{1t}$ and ${\bf q}_{2t}$ are defined in eqs. \ref{qs} and
$\omega_1$ and $\omega_2$ are defined in eq. \ref{omega12}.

If this matrix element was the only one being considered it would be easy to
show that the one-photon exchange with one of the nuclei can be treated in the
equivalent photon approximation. This arises from the structure of the
first term inside the integral in \ref{three2}. This result was 
expected in view
of our results of the last section. But, the two-photon exchange in the lower 
part of the diagram leads to complicated integrals which cannot be simplified
in terms of equivalent photons. 

We thus need to
calculate the six diagrams which are obtained by interchange of the  
two-photon lines, as shown in figure 3, and multiply the 
result by 2 to account for the
same set of diagrams by inverting the roles of each nucleus.
We calculate the amplitudes related to the diagrams by means of the 
same procedures we
adopted in eqs. \ref{M31} through \ref{three2}. We find that he 
matrix elements for the diagrams
(b), (c), (e) and (f) of figure 2 are by a factor $1/\gamma$ smaller 
than those for the 
diagrams (a) and (d).  The amplitude for the diagram (d) in figure 2 yields
\be
{\cal M}_d &=& -{16\over \pi}\ (Z\alpha)^3 
\ {\Psi(0)\over M^{3/2}}  \int d^2 q_t d k_t \ k_t 
\ {
(\omega_2/\omega_1)\ ({\bf q}_{2t} \cdot {\bf e}^*)
\exp\left( {{\bf q}_t \cdot {\bf b}/2} \right)
\over 
\left[ q_{2t}^2+\omega_1^2/\gamma^2\right]
} \nonumber\\
&\times&
\int dk_3 \left[ k_t^2+k_3^2/\gamma^2 \right]^{-1}
\left[ q_{1t}^2+(2\omega_1-\omega_2+k_3)^2/\gamma^2\right]^{-1}
\left[ k_3+\omega_1-\omega_2-k_3^2/2\omega_1\gamma^2\right]^{-1}
\label{three3}
\ee

The corresponding cross section which is obtained from amplitudes
${\cal M}_a$ and ${\cal M}_d$ is given by $d\sigma_a +d\sigma_b
+d\sigma_{int} = 2d\sigma_a$. The interference term $d\sigma_{int}$
yields a contribution of order of $1/\gamma^2$  after azimuthal 
integration, and is disregarded.

The last term in the integrand over $k_3$ dominates the integral
in eq. \ref{three2} and it is strongly peaked (with width of order of
$M/\gamma$) at $k_3 \simeq \omega_1 - \omega_2$. We can replace this
value in the other terms of the integrand and take them out of the
integral. The remaining integral can be done analytically. We get
\be
{\cal M}_a = i 32\ (Z\alpha)^3 
\ {\Psi(0)\over M^{3/2}}  \int d^2 q_t d k_t \ k_t 
\ {
({\bf q}_{1t} \cdot {\bf e}^*)
\exp\left( {{\bf q}_t \cdot {\bf b}/2} \right)
\over 
\left[ q_{1t}^2+\omega_2^2/\gamma^2\right]
} 
{1\over \left[ k_t^2+(\omega_1-\omega_2)^2/\gamma^2 \right]}
{1\over \left[ q_{2t}^2+(2\omega_1-\omega_2)^2/\gamma^2\right]}
\label{three4}
\ee
The same trick can be applied to the amplitude of eq. \ref{three3}.

We now use eq. \ref{delta} and integrate the squared amplitude over $b$,
multiplying by a factor of 2 to account for the amplitude of diagram
(d) of figure 3. 
Again, we insert the nuclear form factors at each of the nuclear vertices to
account for the nuclear sizes. We also change the integration variables
to ${\bf q}_{1t}$ and ${\bf q}_{2t}$.
The final result, after integrating over $\Big|{\bf q}_{1t}\cdot 
\hat{\bf e}^*\Big|^2$, is
\be
{d\sigma \over d P_z} =
1024\pi \ \Big|\Psi(0)\Big|^2
(Z\alpha)^6 { 1\over M^3E}
\  
\int  { dq_{1t} \ q_{1t}^3\ \left[F(q_{1t}^2)\right]^2 \over
\left(q_{1t}^2 + \omega_2^2/\gamma^2 \right)^2}
\ \int  {dq_{2t}\ q_{2t} \ 
\left[F(q_{2t}^2)\right]^2 \over
 \left[ q_{2t}^2+(2\omega_1-\omega_2)^2/\gamma^2\right]^2}
\left[ \int {dk_t \ k_t\ F(k_t^2) \over 
\left(k_t^2+(\omega_1-\omega_2)^2/\gamma^2\right)}
\right]^2
\ee

We now use the relationship between $E$ and $P_z$ to $\omega_1$
and $\omega_2$ and get rid of the meson wavefunction at the origin. 
The wavefunction $\Big|\Psi(0)\Big|^2$ cannot be related to the
$\gamma\gamma$ decay widths. But, vector mesons can decay into
$e^+e^-$ pairs. These decay widths are very well known experimentally.
Following a similar derivation as for the $\gamma\gamma$-decay
the $e^+e^-$ decay-width of the vector mesons can be shown \cite{RW67}
to be equal to $\Gamma_{e^+e^-}=16\pi \alpha^2 \Big|\Psi(0)\Big|^2/3M^2
\ (3\cdot \sum_i Q_i^2)$. Inserting these results in the above equation,
the factor $(3\cdot \sum_i Q_i^2)$ will cancel out for the same
reason as explained in section 3, and we get
\be
{d\sigma \over d\omega} =  \sigma^{(-)} \
\ {n(\omega)\over \omega}\ \ H (M,\omega)
\ee
where
\be
 \sigma^{(-)}=96\pi
{\Gamma_{e^+e^-}\over M^3} \ ,
\ee
with $n(\omega)$ given by  \ref{epaf2} and
\be
H(M,\omega) = Z^4\alpha^3 M^2
\int 
{dq_{2t}\ q_{2t} \ \left[F(q_{2t}^2)\right]^2
\over
\left[ q_{2t}^2+(M^2/2\omega-\omega)^2/\gamma^2\right]^2}
\left[ \int {dk_t \ k_t\ F(k_t^2) \over
\left(k_t^2+(M^2/4\omega-\omega)^2/\gamma^2\right)^2}\right]^2
\label{three5}
\ee 

The above formulas should also be valid for the production of the
ortho-positronium in ultra-peripheral collisions of relativistic
heavy ions. For RHIC ($Au+Au$) we obtain $\sigma = 11.2$ mb, while
for the LHC ($Pb+Pb$) we get $\sigma = 35$ mb. These numbers are
somewhat larger, but are in the same ballpark as the results given
in ref. \cite{serbo}. Note that
Coulomb corrections \cite{serbo} substantially modify the positronium production 
cross section in relativistic heavy ion collisions. This is not considered
in the present approach.   

In table 2 we present the cross sections for the production of
vector mesons by means of the three-photon fusion process. We use the
form factor given by eq. \ref{formfact}.

\bigskip

\begin{center}
Table 2. {\small Cross sections for three-photon production of vector ($C=odd$) mesons 
at RHIC ($Au+Au$) and at LHC ($Pb+Pb$).}

\bigskip
\begin{tabular}{l|c|c|c|c|c|c|r}
\hline
& meson     & mass  [MeV]& $\Gamma_{e^+e^-}$ [keV]& 
$\sigma^{(-)}$ [nb]
&$\sigma^{RHIC}$ [nb] & $\sigma^{LHC}$ [nb] \cr      
\hline
& $\rho^0$   &  770  & 6.77 &1740       & 137   & 1801    \cr
& $\omega$    &   782  & 0.60     & 147 &  13  &  163 \cr
& $J/\psi$& 3097 & 5.26     &  21     &  31  & 423 \cr
& $\Psi$'& 3686 & 2.12     &  5       &  12  & 155 \cr
\hline
\end{tabular}

\end{center}

\bigskip

We see that the cross sections for the production of vector mesons
in ultra-peripheral collisions  of relativistic heavy ions are small. 
They do not compare to the production of vector mesons in 
central collisions. In principle, one would expect that the cross sections
for three-photon production would scale as $(Z\alpha)^3$, which is
an extra $Z\alpha$ factor compared to the two-photon fusion
cross sections. However, the integral over the additional photon momentum
decreases the cross section by several orders of magnitude.  

\section{Conclusions}
We have carried out a derivation of the production of mesons in
ultra-peripheral collisions of relativistic heavy ions in terms of a
projection procedure. This is useful in order to study the
virtuality content of the exchanged photons.
We have shown that the cross section
for the production of the (para-)positronium is 
the same as that obtained by another calculation \cite{serbo}.

It has also been shown that the inclusion of nuclear form factors 
leads to cross sections for two-photon fusion which agree with those
obtained by the equivalent photon approximation. As a byproduct we
extended the calculation to the production of vector mesons. We have
shown that their cross sections are very small and can be neglected
for practical purposes.

\bigskip\bigskip

This work has been authored under Contract No. DE-AC02-98CH10886 
with the U.S. Department of Energy, and partial
support from the Brazilian funding agency 
MCT/FINEP/CNPQ(PRONEX), under contract No. 41.96.0886.00, is 
also acknowledged. One of us (CAB) is a fellow of the  John Simon Guggenheim 
Foundation. We had also financial support from FAPESP under
contract 1999/12987-5.

\end{document}